\documentclass[aps,prb,reprint,groupedaddress]{revtex4-1}
\usepackage{graphicx}
\usepackage{xspace}
\usepackage{amssymb}
\usepackage{dcolumn}
\newcommand{\baco}{BaCo$_{\text{2}}$V$_{\text{2}}$O$_{\text{8}}$\xspace}
\newcommand{\srco}{SrCo$_{\text{2}}$V$_{\text{2}}$O$_{\text{8}}$\xspace}
\newcommand{\basrco}{Ba$_{\text{1-x}}$Sr$_{\text{x}}$Co$_{\text{2}}$V$_{\text{2}}$O$_{\text{8}}$\xspace}
\newcommand{\bacom}{Ba(Co$_{0.95}M_{0.05}$)$_2$V$_2$O$_8$\xspace}
\newcommand{\amo}{$AM_{\text{2}}$V$_{\text{2}}$O$_{\text{8}}$\xspace}

\newcommand{\SrNi}{SrNi$_{\text{2}}$V$_{\text{2}}$O$_{\text{8}}$\xspace}
\newcommand{\BaCu}{BaCu$_{\text{2}}$V$_{\text{2}}$O$_{\text{8}}$\xspace}
\newcommand{\SrMn}{SrMn$_{\text{2}}$V$_{\text{2}}$O$_{\text{8}}$\xspace}

\newcommand{\bacomg}{Ba(Co$_{0.95}$Mg$_{0.05}$)$_2$V$_2$O$_8$\xspace}
\newcommand{\basrA}{Ba$_{0.9}$Sr$_{0.1}$Co$_2$V$_2$O$_8$\xspace}

\newcommand{\basrx}{Ba$_{1-x}$Sr$_{x}$Co$_2$V$_2$O$_8$\xspace}
\newcommand{\tn}{$T_{\text{N}}$\xspace}

\begin{document}

\title{Substitution effects on the temperature vs. magnetic-field phase diagrams of the quasi-1D effective Ising spin-1/2 chain system BaCo$_2$V$_2$O$_8$}
\author{S.~K.~Niesen$^1$}
\author{O.~Breunig$^1$}
\author{S.~Salm$^1$}
\author{M.~Seher$^1$}
\author{M.~Valldor$^{1,2}$}
\author{T.~Lorenz$^1$}
\email[]{tl@ph2.uni-koeln.de}
\affiliation{$^1$II. Physikalisches Institut, University of Cologne, Z\"ulpicher Str. 77, 50937 K\"oln, Germany\\
                 $^2$Max-Planck-Institut f\"ur Chemische Physik fester Stoffe, N\"othnitzer Str. 40, 01187 Dreden, Germany}

\date{\today}

\begin{abstract}
\baco is a one-dimensional antiferromagnetic spin-1/2 chain system with pronounced Ising anisotropy of the magnetic exchange. Due to finite interchain interactions long-range antiferromagnetic order develops below $T_{\rm N} \simeq 5.5$~K, which is accompanied by a structural distortion in order to lift magnetic frustration effects. The corresponding temperature $vs. $ magnetic-field phase diagram is highly anisotropic with respect to the magnetic-field direction and various details are still under vivid discussion. Here, we report the influence of several substitutions on the magnetic properties and the phase diagrams of \baco . We investigate the  substitution series Ba$_{\text{1-x}}$Sr$_{\text{x}}$Co$_{\text{2}}$V$_{\text{2}}$O$_{\text{8}}$ over the full range $0\le x \le 1$ as well as the influence of a partial substitution of the magnetic Co$^{2+}$ by small amounts of other magnetic transition metals or by non-magnetic magnesium. In all cases, the phase diagrams were obtained on single crystals from magnetization data and/or high-resolution studies of the thermal expansion and magnetostriction.
\end{abstract}

\maketitle

\section{Introduction}
\label{sec:Intro}
\baco is one representative of the material class \amo with earth-alkaline metals  $A=$~Sr, Ba and different divalent transition metals on the $M$ site. Commonly, these compounds have a similar tetragonal crystal structure, which contains screw chains of $M$O$_6$ octahedra running along the $c$ axis. These screw chains are separated by the non-magnetic $A^{2+}$ ions and VO$_4$ tetrahedra (with nonmagnetic V$^{5+}$). Depending on the transition metal $M=$~Cu, Ni, Co, or Mn, quasi-one-dimensional (1D) spin systems with different spin quantum numbers are realized, {\it e.g.}, dimerized $S=1/2$ Heisenberg chains for \BaCu~\cite{He2004,Note1}, $S=1$ Haldane chains for \SrNi~\cite{Bera2013}, or $S=5/2$ Heisenberg chains for \SrMn~\cite{Niesen2011}.
For Co$^{2+}$ with $3d^7$, the partial occupation of the $t_{2g}$ orbitals results in an effective orbital moment $\tilde{\ell}=1$ and the spin-orbit coupling between $\tilde{\ell}$ and the total spin $S=3/2$ is typically in the same energy range as the crystal-field splitting of the $t_{2g}$ orbitals due to a distortion of the CoO$_6$ octahedra. As a consequence, the Co$^{2+}$ magnetic moments are often strongly anisotropic~\cite{Goodenough1968, Csiszar2005}. 
 For example, an easy-axis (or Ising) anisotropy is observed in K$_2$CoF$_4$ with compressed CoF$_6$ octahedra~\cite{Folen1968}, while an easy-plane (or XY) anisotropy is present in La$_{2-x}$Sr$_x$CoO$_4$ with elongated CoO$_6$ octahedra~\cite{Hollmann2008}.
In the case of \baco, the CoO$_6$ octahedra are significantly compressed along the $c$ axis and a strong Ising anisotropy with $c$ as the magnetic easy axis is reported~\cite{He2005A,He2005}.
Consequently, \baco represents a quasi-1D effective Ising $S=1/2$ chain described by the $XXZ$ hamiltonian
\begin{equation}
\mathcal{H}= \sum_i J\left\{S_i^z S_{i+1}^z + \varepsilon \left(S_i^x S_{i+1}^x+S_i^y S_{i+1}^y  \right) \right\} +   g \mu_{\text{B}} {\bf S}_i {\bf H} \, , 
\label{hamil}
\end{equation}
where $J/k_{\text{B}}\simeq 65$~K, $\varepsilon \simeq 0.46$ and the anisotropic $g$ factors $g^{\|c}\simeq 6.3$ and $g^{\perp c}\simeq 3.2$ have been derived from an analysis of the low-temperature magnetization curves~\cite{Kimura2006}. 
At $T_{\text{N}}\simeq 5.5$~K, \baco orders antiferromagnetically due to the presence of finite interchain couplings. 
As expected from the Ising anisotropy, the ordered moments point along the $c$ direction. 
Within the chains ($i.e. \| c$) nearest-neighbor (NN) spins are antiparallel to each other, but perpendicular to the chains, \textit{i.e.} within the (001) planes, NN spins point either in the same or in the opposite direction~\cite{Kimura2008A, Kawasaki2011,Grenier2013}. 
Consequently, the ordered phase typically contains different magnetic domains, which are rotated by 90$^\circ$ and translated by $c/4$ with respect to each other~\cite{Grenier2013}. Recently, we found that the antiferromagnetic ordering of \baco is accompanied by a small orthorhombic distortion within the $ab$ planes, which can be traced back to a finite magneto\-elastic coupling of the interchain couplings~\cite{Niesen2013}. Thus, magnetic and structural domains are coupled to each other and, in fact, the under ambient conditions twinned crystals of \baco can be, at least partially, detwinned either by cooling under uniaxial pressure or by applying a magnetic field along the [100] direction~\cite{Niesen2013}. 

A further consequence of the Ising anisotropy in \baco is its strongly anisotropic temperature $vs.$ magnetic-field phase diagram~\cite{He2005, He2006A, Kimura2008, Kimura2008A, Kimura2009, Kimura2010,Yamaguchi2011,Kawasaki2011,Klanjsek2012,Zhao2012,Grenier2013,Niesen2013}. 
Magnetic fields along the easy axis, {\it i.e.} $H\| c$, strongly suppress \tn and above about 4~T an incommensurate (IC) magnetic order is observed with a weakly field-dependent transition temperature of $T_{\text{IC}}\approx  1$~K. Within this IC phase, there are some experimental indications for the occurrence of additional phase transitions~\cite{Yamaguchi2011, Klanjsek2012}, but this issue is not settled yet. Concerning the field direction $H\perp c$, a very weak decrease of  $\partial T_{\text{N}}/\partial H \approx -0.1$~K/T has been originally reported for the field range up to 9~T~\cite{He2006A}, which roughly corresponds to the observed saturation field of about 40~T measured at $T=1.3$~K~\cite{Kimura2006}. 

Recently, however, it has been realized that this very weak field influence is only present for $H\| [110]$~\cite{Kimura2013, Niesen2013}. 
Rotating the field by $45^\circ$ within the $ab$ plane,  {\it i.e.} a field $H\| a$, causes a much stronger decrease of \tn and it is completely suppressed above about 10~T. As was shown in Ref.~\onlinecite{Kimura2013}, such an in-plane anisotropy can be explained by a small tilt ($\sim 5^\circ$) of the local easy axis of the CoO$_6$ octahedra with respect to the $c$ axis. 
Due to the $4_1$ screw axis along $c$, the direction of the local easy axis changes with a four-step periodicity between neighboring Co ions along $c$. 
As a consequence, applied external magnetic fields cause staggered effective fields in the transverse directions. 
Here, $H \| a$ results in effective fields $h_b\|b$ and $h_c\|c$ with periodicities $+-+-$ and $+--+$, respectively, while $H \| [110]$ only causes an effective field $h_c\|c$ with periodicity $+0-0$ and leaves $h_b=0$. According to Ref.~\onlinecite{Kimura2013}, the presence or absence of a staggered field $h_b\|b$ is the origin of the above-described strong in-plane anisotropy of $T_{\rm N}(H)$ of \baco for $H\perp c$.       

As the direction of the local easy axis is related to structural distortions, we studied the influence of a partial substitution of the Ba$^{2+}$ ions by smaller Sr$^{2+}$ ions on the temperature $vs.$ magnetic-field phase diagram. A further motivation to investigate the substitution series \basrco was the fact that there are only few studies on \srco ~\cite{He2006Sr, He2007Sr, Lejay2011}. The crystal structures of both end members are very similar, but \baco is centrosymmetric (space group $I4_1/acd$ No.~142)~\cite{Wichmann86}, while \srco lacks centrosymmetry (space group $I4_1cd$ No.~110)~\cite{Osterloh94, Note2}. 
 
Concerning the magnetic properties of \srco, the magnetic susceptibility in the high-temperature paramagnetic phase has a similar Ising anisotropy to that of \baco. Moreover, \srco also shows N\'{e}el order at an only weakly reduced $T_{\rm N}\simeq 4.5$~K. Despite these similarities, it was concluded that the temperature $vs.$ magnetic-field phase diagrams of \baco and \srco would be drastically different, because in \srco a two-step transition in low fields and a rather moderate anisotropy of $T_{\rm N}(H)$ between $H || c$ and $H \perp c$ had been observed in Refs.~\onlinecite{He2006Sr, He2007Sr}. However, such a two-step transition was not confirmed in a more recent study~\citep{Lejay2011} and, as will be shown below, the seemingly very different anisotropies of $T_{\rm N}(H)$ reported for the two end members of \basrco ~\cite{He2006A, He2006Sr} essentially result from the in-plane anisotropy of $T_{\rm N}(H\perp c)$, which has been discovered only recently \cite{Kimura2013, Niesen2013}. This in-plane anisotropy is present for all $x$ and systematically increases as a function of increasing Sr content. In addition, we also studied the influence of in-chain substitutions by partially replacing the effective Ising $S=1/2$ Co$^{2+}$ ion by transition metal ions with different spin quantum numbers (Cu$^{2+}$, Ni$^{2+}$, Mn$^{2+}$) or by nonmagnetic Mg$^{2+}$. The resulting change of \tn in zero magnetic field is rather different, but in all cases the magnetic anisotropies are essentially preserved. 

\section{Crystal growth and characterization}

\begin{figure}[t]
	\centering
		\includegraphics[width=0.95\linewidth]{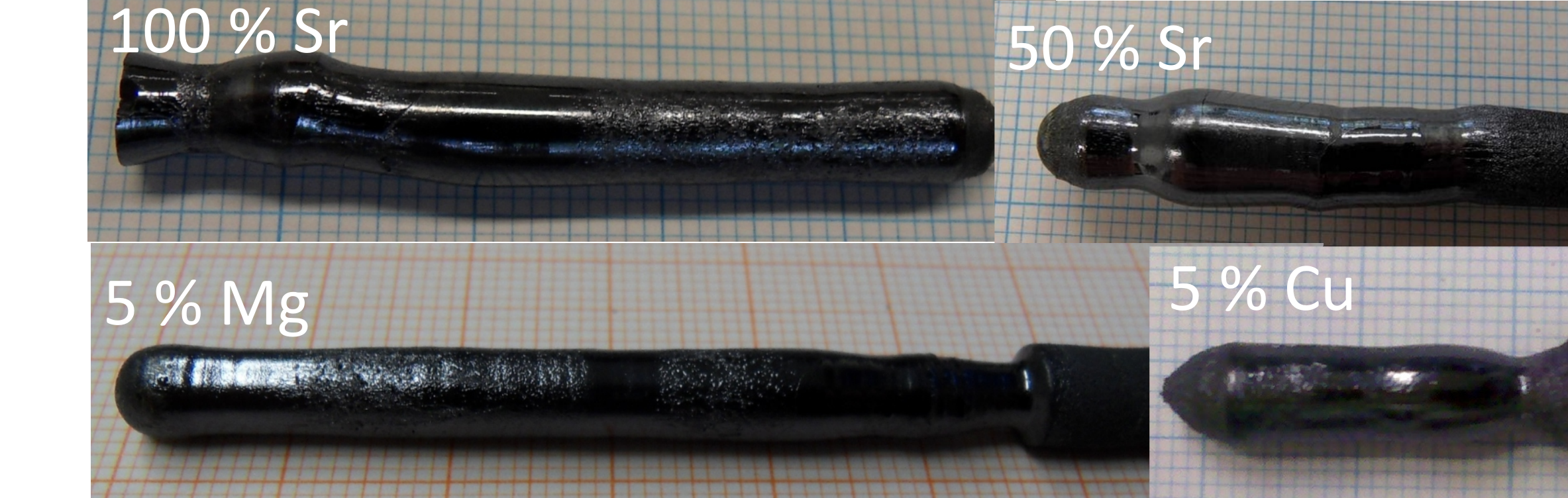}
		\includegraphics[width=1.00\linewidth]{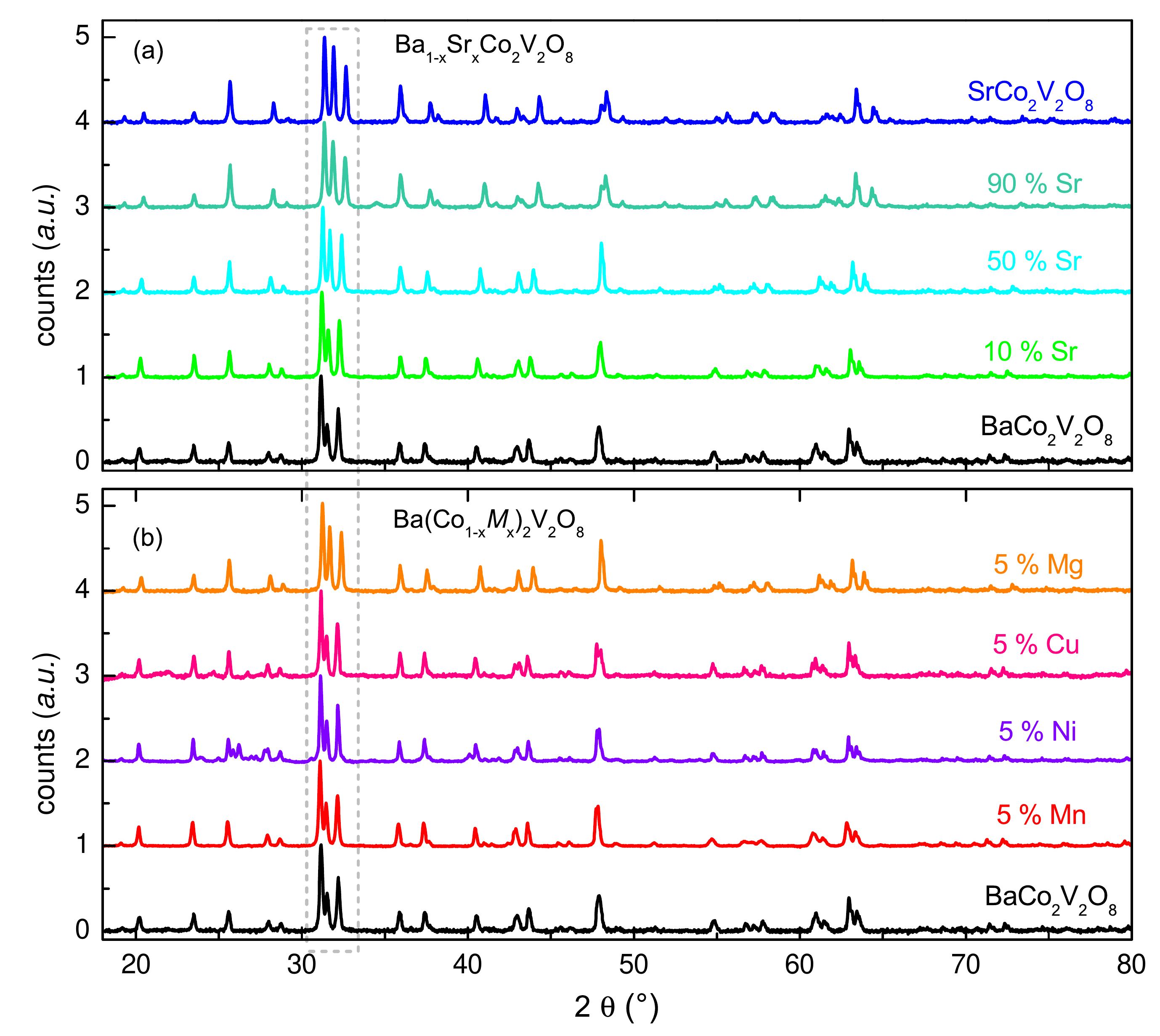}
		\caption{(color online) Single crystals of Ba$_{1-x}$Sr$_x$Co$_2$V$_2$O$_8$ and of \bacom with different $x$ or $M$, respectively, and X-ray powder diffraction patterns of (a) Ba$_{1-x}$Sr$_x$Co$_2$V$_2$O$_8$ and (b) \bacom. The data are normalized to the maximum intensity and are shifted with respect to each other for clarity.}
	\label{fig:Samples-powder}
\end{figure}
All measurements were performed on single crystals synthesized by the floating-zone method in a four-mirror image furnace (FZ-T-10000-H-VI-VP, Crystal Systems Inc.). In general, the procedure was done in close analogy to the growth of \baco described in Refs.~\onlinecite{Lejay2011, Niesen2013}. 
For the Sr-substituted series, appropriate amounts of BaCoO$_3$ (99+\% Merck) and SrCoO$_3$ (99.99\% Alfa Aesar) were mixed with Co$_3$O$_4$ (99.5\% Alfa Aesar), V$_2$O$_5$ (99.5\% Strem Chemicals) and polycrystalline feed and seed rods of \basrco were prepared in a solid-state reaction. For the growth of \bacom the appropriate mixtures of Co$_3$O$_4$ with either CuO (99.995\%, Alfa Aesar),  NiO (99.998\%, Alfa Aesar) or MnO (99\%, Aldrich) were used for $M={\rm Cu}$, Ni or Mn, respectively. Due to the high stability of MgO, the analogous route did not work for the synthesis of \bacomg and this crystal was therefore grown from a mixture of Ba(OH)$_2 \cdot$ 8~H$_2$O ($\geq$ 98\% Sigma Aldrich), Mg(OH)$_2$ (95+\% Alfa Aesar), Co$_3$O$_4$, and V$_2$O$_5$. For all compositions, the single-crystal growth in the image furnace was done in air at ambient pressure and performed by the so-called twice-scanning method. The first run has been done rather fast with growth rates between 2 and 5~mm/h, while the second run was performed with typical growth rates of $\approx 0.5$~mm/h. 

\begin{figure}[t]
	\centering
		\includegraphics[width=1.00\linewidth]{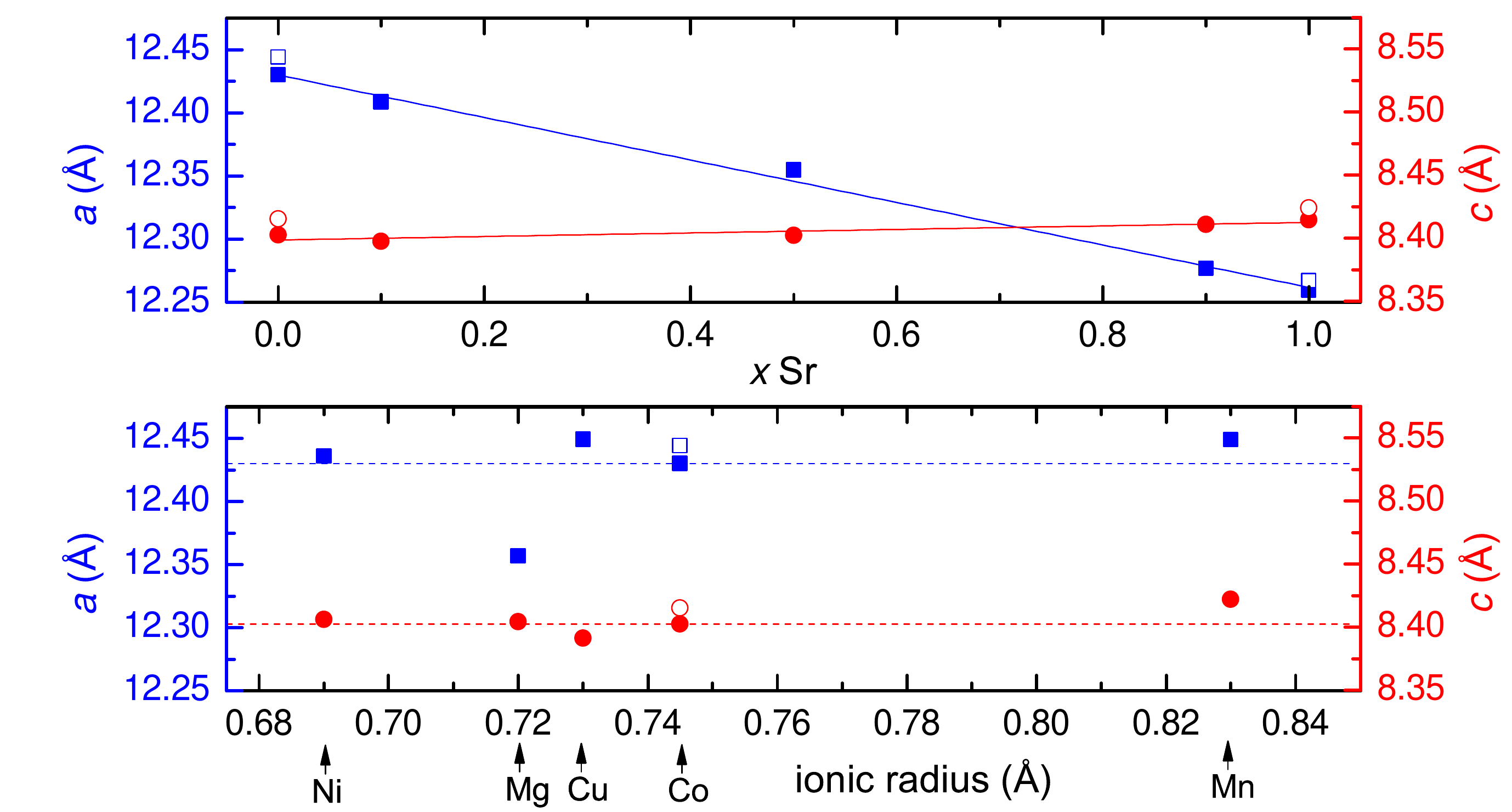}
	\caption{(color online) Lattice constants $a$ (left axes) and $c$ (right axes) of Ba$_{1-x}$Sr$_x$Co$_2$V$_2$O$_8$ as a function of $x$ (top) and of \bacom as a function of the ionic radius of $M$ (bottom); open symbols are from Refs. \onlinecite{Wichmann86, Osterloh94}. The solid lines are linear fits of $a$ and $c$ as functions of the Sr content and the dashed lines in the lower panel mark the lattice constants of \baco.}
	\label{fig:GKs}
\end{figure}

In order to check phase purity, small parts of the crystals were crushed and studied by X-ray powder diffraction in Bragg-Brentano geometry in a D5000 Stoe diffractometer with CuK$_{\alpha_{1,2}}$ radiation and a position-sensitive detector. 
The refinement was done with Fullprof~\cite{RodriguezCarvajal199355} using either the tetragonal space group $I4_1$/$acd$ or $I4_1cd$ reported for \baco or \srco, respectively~\cite{Wichmann86, Osterloh94}. From powder-diffraction data it is not possible to unambiguously distinguish these space groups, but the three largest peaks can be taken as an indicator which structure is present (see dashed box of Fig. \ref{fig:Samples-powder}). For $I4_1$/$acd$, the central peak is the weakest of the three, but for $I4_1cd$ it is the right one. With increasing Sr substitution, the central peak's intensity increases as may be expected from the different structures of  the end members, but we cannot judge where the space group changes from centro- to non-centrosymmetric. For the analysis, the centrosymmetric space group has been chosen up to the Sr substitution of 10\% and for all \bacom crystals, while  $I4_1cd$ has been used otherwise.
In Fig.~\ref{fig:GKs} we show the resulting lattice constants $a$ and $c$ for the Ba- and Co-substituted crystals. For comparison, the literature data of \baco and \srco are also shown. With increasing Sr content, we find a significant, almost linear decrease for the $a$ axis and a weak increase of the $c$ axis. As might be expected from the rather low concentrations, the partial substitution of Co by other transition metal ions leaves $a$ and $c$ almost unchanged. The 5\% Mg substitution, however, causes a significant reduction of $a$ despite the fact that its tabulated ionic radius is close to those of the transition metals. This is naturally interpreted by a different  bonding character of the Mg--O bonds compared to the $M$--O bonds.

\section{Low-field measurements}

\begin{figure}[t]
	\centering
		\includegraphics[width=\linewidth]{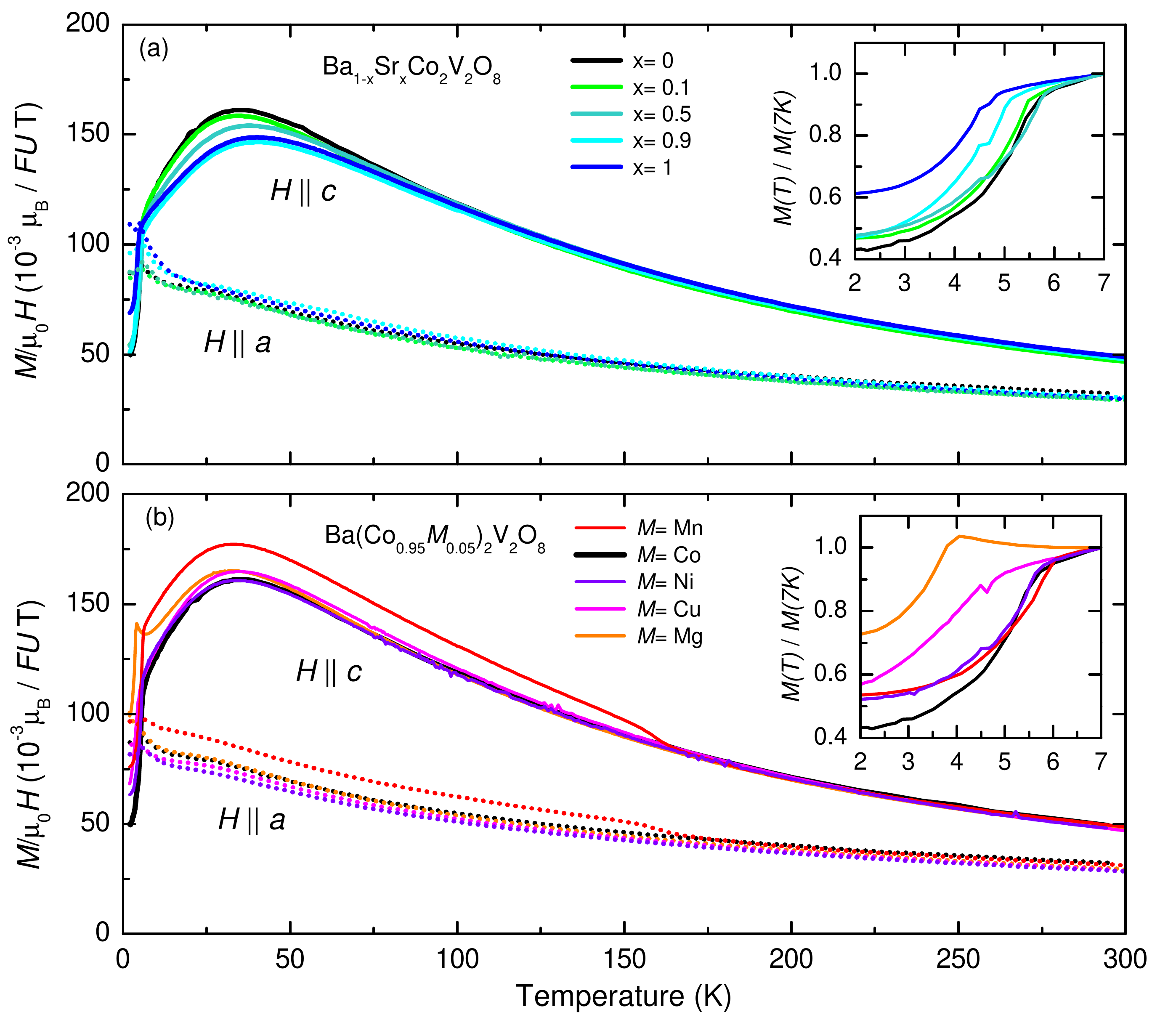}
	\caption{(color online) Anisotropic susceptibility of Ba$_{1-x}$Sr$_x$Co$_2$V$_2$O$_8$  (top) and \bacom (bottom) measured in a magnetic field of $H=100$~mT applied either along the easy axis $c$ (solid lines) or along $a$ (dashed lines). The insets display an enlargement of the normalized data $M(T)/M(7\, \rm{K})$ for $H || c$ in the low-temperature range. The additional anomaly around 160~K in the Mn-substituted sample probably arises from a ferromagnetic impurity.}  
	\label{fig:susc}
\end{figure}

The magnetic susceptibility data $\chi^{i}(T)$ were measured for different field directions $i$ on both substitution series from about 2~K to 300~K using a commercial SQUID magnetometer (MPMS, Quantum Design Inc.). 
The resulting low-field ($H=0.1$~T) data  are summarized in Fig.~\ref{fig:susc}. 
The above-mentioned strong Ising anisotropy in the paramagnetic high-temperature phase of \baco, that is $\chi^{H\|c}/\chi^{H\perp c} \approx 1.5$--2 for $T> 50 $~K, remains essentially preserved upon the full range of Sr substitution as well as for the partial substitution on the Co$^{2+}$ sites. In this low-field high-temperature range no in-plane anisotropy between $H|| a$ and $H || [110]$ is present, neither in any of the compounds studied here nor in the mother compound~\cite{Niesen2013}.
Around 5~K, there are clear anomalies in the susceptibility data of all samples for both field directions, which signal the long-range antiferromagnetic ordering. 
The insets of Fig.~\ref{fig:susc} display the normalized magnetization data $M(T)/M(7~{\rm K})$ for $H\|c$ of both substitution series. Over the entire range of Sr substitution we observe a weak decrease of $\Delta T_{\rm N} \simeq 1$~K. As a criterion of \tn, we use the maxima of the temperature derivatives $\partial \chi^{i}/\partial T$. From the magnetization data of both field directions together with the peak positions of the thermal-expansion measurements $\alpha_i =1/L_{i0}\, \partial L_{i}/\partial T$ measured along different lattice directions $i=c$, $a$, or [110] (see below) in zero magnetic field, we find that \tn remains practically unchanged from $0\le x\le 0.5$ and then decreases to $T_{\rm N} \simeq 4.5$~K for pure \srco , see table~\ref{tab:crit}. 
Concerning the Co substitution, we observe no change or even a small increase of \tn for $M={\rm Ni}$ or Mn, while 5\% of Cu or Mg reduce \tn to $\simeq 4.5$~K or $\simeq 3.7$~K, respectively.   

\begin{table}
	\centering
	\caption{Critical temperatures \tn in zero magnetic field and critical magnetic fields obtained by extrapolating $H_{crit}^{\| a,c}(T)$ to $T\rightarrow 0$. Note that  for $H\| c$ and $H\| a$ different phases are realized above the respective critical fields. The corresponding ratios $H_{crit}^{\| a,c}(T\rightarrow 0)/T_{\rm N}(H=0)$ are displayed in Fig.~\ref{fig:rel}. }
		\begin{tabular}{rrccccrcccc}
				\hline
		 & \multicolumn{5}{c}{\basrco} & \multicolumn{5}{c}{\bacom}		\\
	& $x=0$ & 0.1 & 0.5& 0.9 & 1 &  $M=$ & Mn & Ni & Cu &  Mg \\
\hline
$T_{\rm N}$(K) &   5.5& 5.5& 5.5 & 4.9 & 4.5 &  & 5.8 & 5.5 & 4.4 & 3.7 \\
$H_{crit}^{\| c}$(T) 	& 3.9 & 3.9 &  3.9 & 3.7 & 3.5 & &3.1&3.7 &3.0  &1.6\\
$H_{crit}^{\| a}$(T)&  9.7 & 8.7 &8.0& 6.4  &5.3  & &12&9.5& 8.2 &7.2\\
\hline
		\end{tabular}
		\label{tab:crit}
\end{table}

The different influence of the various substitutions on the antiferromagnetic ordering can be understood qualitatively as follows. The three-dimensional ordering of Ising (as well as, {\it e.g.}, Heisenberg or XXZ) spin chains requires a finite interchain coupling, but the actual value of \tn depends only weakly on the interchain coupling~\cite{Imry1975,Yurishchev1991}. 
Because the Sr substitution mainly changes the $a$ axis, {\it i.e.} the interchain distance (see Fig.~\ref{fig:GKs}), one may expect a stronger influence on the interchain couplings, while the intrachain coupling should remain essentially unchanged. This is supported by the observation that the anisotropic magnetic susceptibility is essentially preserved over the entire \basrco series in the temperature range well above \tn , where the interchain couplings are of minor importance. 

Concerning the in-chain substitutions, one can also expect little changes of $\chi^i(T)$ in the higher temperature range, where the correlation length $\xi(T)$ is significantly shorter than the average distance of 20 lattice sites for the 5\% substitution level. However, a significant influence is expected for \tn, because the spin chains are cut into finite chain segments and thus the low-temperature increase of $\xi(T)$ is limited, if the intrachain magnetic coupling via the substituted $M$ ion is effectively blocked~\citep{Imry1975impurity,Eggert2002}. The degree of (de-)coupling of the finite chain segments depends on the magnetic properties of the substituting ion $M$, which obviously influences the NN intrachain exchange. 
In addition, the next nearest neighbor (NNN) intrachain exchange contributes and an effective coupling of the finite chain segments via the neighboring chains due to the finite interchain coupling might be important, too. Because the nonmagnetic Mg causes a strong suppression of \tn, while the magnetic Mn and Ni ions have almost no influence we conclude that in \baco the NN interaction is the dominant coupling mechanism of the finite chain segments. Moreover, the intermediate suppression of \tn for the Cu substitution suggests that the coupling via this $S=1/2$ ion is already considerably reduced whereas there seems to be little change when the $S=3/2$ Co$^{2+}$ ions are replaced by  $S=1$ Ni$^{2+}$ or  $S=5/2$ Mn$^{2+}$. We think that this difference results from the fact that Co$^{2+}$,  Ni$^{2+}$, and Mn$^{2+}$ have half-filled $e_g$ orbitals, whereas there is only a single hole in the $e_g$ orbital of Cu$^{2+}$. Furthermore, the quantum fluctuations are significantly larger for  $S=1/2$ than for higher spin quantum numbers. 

Because the interchain distance considerably decreases with increasing Sr content, one may expect increasing interchain exchange couplings, which seems to be in conflict with the observed weakly decreasing \tn. As mentioned above, however, in the ordered phase of \baco NN spins in the (001) planes point either parallel or antiparallel to each other, that is, the symmetry of the magnetic structure is less than tetragonal~\cite{Kawasaki2011, Grenier2013}. As we have discussed in more detail in Ref.~\onlinecite{Niesen2013}, the magnetic structure of \baco typically arises when there is a frustration between NN ($J^\perp_{NN}$) and NNN interchain interactions ($J^\perp_{NNN}$). Thus, our data suggest that the decreasing interchain distance in the \basrco series changes $J^\perp_{NN}$ and $J^\perp_{NNN}$ in a way that the resulting frustration increases, which in turn decreases the effective interchain interaction and therefore causes a weakly decreasing \tn with increasing $x$. 

Neutron scattering data of the magnetic structure are available only for \baco ~\cite{Kawasaki2011, Grenier2013}. For \srco, a different magnetic structure with a non-collinear spin arrangement was proposed~\cite{He2006Sr,He2007Sr}. However, this suggestion is based on a very limited set of magnetization data and was questioned by a more recent magnetization study~\cite{Lejay2011}. One characteristic feature of the magnetic structure of \baco is the presence of 2 types of antiferromagnetic domains and when cooling under ambient conditions the crystals are heavily twinned. As we have shown in Ref.~\onlinecite{Niesen2013}, the magnetoelastic coupling of the in-plane couplings $J^\perp_{NN}$ and $J^\perp_{NNN}$ causes a small orthorhombic splitting $a_o> a > b_o$. Here, $a_o$ and $b_o$ denote the lattice constants in the orthorhombic phase below \tn, while $a$ refers to the tetragonal paramagnetic phase. The orthorhombic splitting is too small to be seen in standard diffraction data and the twinning complicates its detection in macroscopic data. However, the degree of twinning can be influenced by cooling a crystal under different uniaxial pressures $p_i\|a$, which then results in a pressure-dependent overall spontaneous strain for $T<T_{\rm N}$ that can be clearly seen in high-resolution measurements of the macroscopic thermal expansion along the tetragonal $a$ axis~\cite{Niesen2013}.

\begin{figure}[t]
	\centering
		\includegraphics[width=1.00\linewidth]{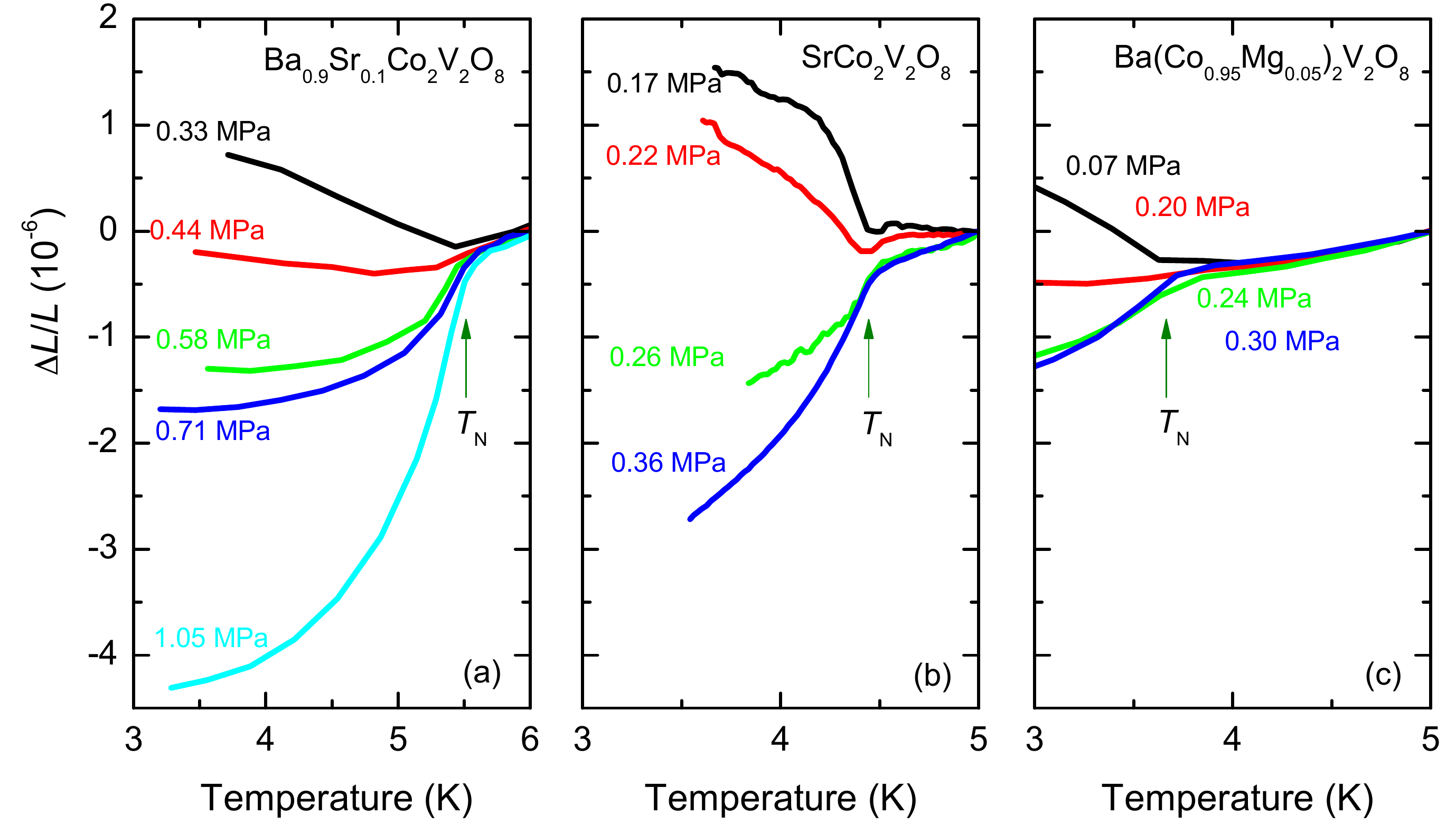}
	\caption{(color online) Zero-field thermal expansion of the tetragonal $a$ axis of  \basrco for $x=0.1$ and $x=1$ and of Ba(Co$_{0.95}$Mg$_{0.05}$)$_2$V$_2$O$_8$ measured for different uniaxial pressures $p_i || a$. The length changes $\Delta L = L(T)- L_0$ are related to $L_0$ at the respective maximum temperature. The pressure dependencies of $\Delta L$ below \tn arise from a partial detwinning with increasing $p_i$ (see text and Ref.~\onlinecite{Niesen2013}).}
	\label{fig:tadnull}
\end{figure}

Fig.~\ref{fig:tadnull} displays the relative length changes $\Delta L(T)/L_0$ measured along  $a$ on \basrA , \srco, and \bacomg by using a home-built capacitance dilatometer~\cite{Pott1977}.
During the measurement, the crystal is clamped  via two CuBe springs that fix the moveable capacitor plate and the corresponding uniaxial pressure $p_a$ can be varied by varying the base capacitance. For the actual  dilatometer, uniaxial pressures are in the MPa range for samples with typical cross sections of a few mm$^2$. As can be seen in Fig.~\ref{fig:tadnull}, below \tn all three crystals evolve a finite macroscopic spontaneous strain, whose magnitude and sign can be systematically changed by increasing the uniaxial pressure. As discussed in more detail in Ref.~\onlinecite{Niesen2013} for the parent compound \baco, this pressure dependence clearly reveals that the antiferromagnetic ordering is accompanied by an orthorhombic splitting and that the crystals are twinned below \tn . The fact that this characteristic feature of the zero-field magnetic structure of \baco remains preserved over the full substitution range of Ba by Sr as well as for the partial substitution of Co by Mg strongly suggests that the magnetic structure of all these crystals does not change significantly. As will be seen below, this conclusion is further supported by the temperature {\it vs.} magnetic-field phase diagrams discussed in the next section. Nevertheless, a definite proof of this claim requires a microscopic determination of the magnetic structure.

\section{High-field measurements}

The phase diagram of \baco reflects the expected Ising anisotropy which arises from the fact that a magnetic field $H\perp c$ is symmetry breaking, while $H\| c$ is not. 
In addition, there is an in-plane anisotropy with respect to $H\| [100]$ and $H\| [110]$~\cite{Niesen2013,Kimura2013}. 
Therefore, the temperature {\it vs.} magnetic-field phase diagrams of the present materials for these three field directions will be discussed separately. In most cases, the phase boundaries were determined from the extrema of the derivatives of the thermal-expansion and magnetostriction data $\Delta L_i(T,H)$, which were measured on a capacitance dilatometer in the temperature range from about 0.3 to 10~K in magnetic fields up to 14~T in a longitudinal configuration, {\it i.e.} with $\Delta L_i(T,H)\| H$. 
For those crystals (or field directions), where $\Delta L_i(T,H)$ was not studied, the phase boundaries are determined from magnetization data $M(T,H)$ measured in the SQUID magnetometer down to 2~K in magnetic fields up to 7~T.~\cite{Note3} 

The data presentation in the following subsections is restricted  to characteristic measurements of $\Delta L_i(T,H)$ on three representative crystals, which cover the entire substitution range. \basrA is very close to the parent compound \baco that has been discussed in detail in Ref.~\onlinecite{Niesen2013}, \srco is the other end member of the out-of-chain substitution series and \bacomg shows the most severe influence of the in-chain substituted crystals. The $\Delta L_i(T,H)$ data obtained on all the other crystals as well as the magnetization data will not be shown, but the corresponding temperature {\it vs.} magnetic-field phase diagrams will be discussed for all compositions.

\subsection{Magnetic field parallel to [001]}

\begin{figure}[t]
	\centering
		\includegraphics[width=1.00\linewidth]{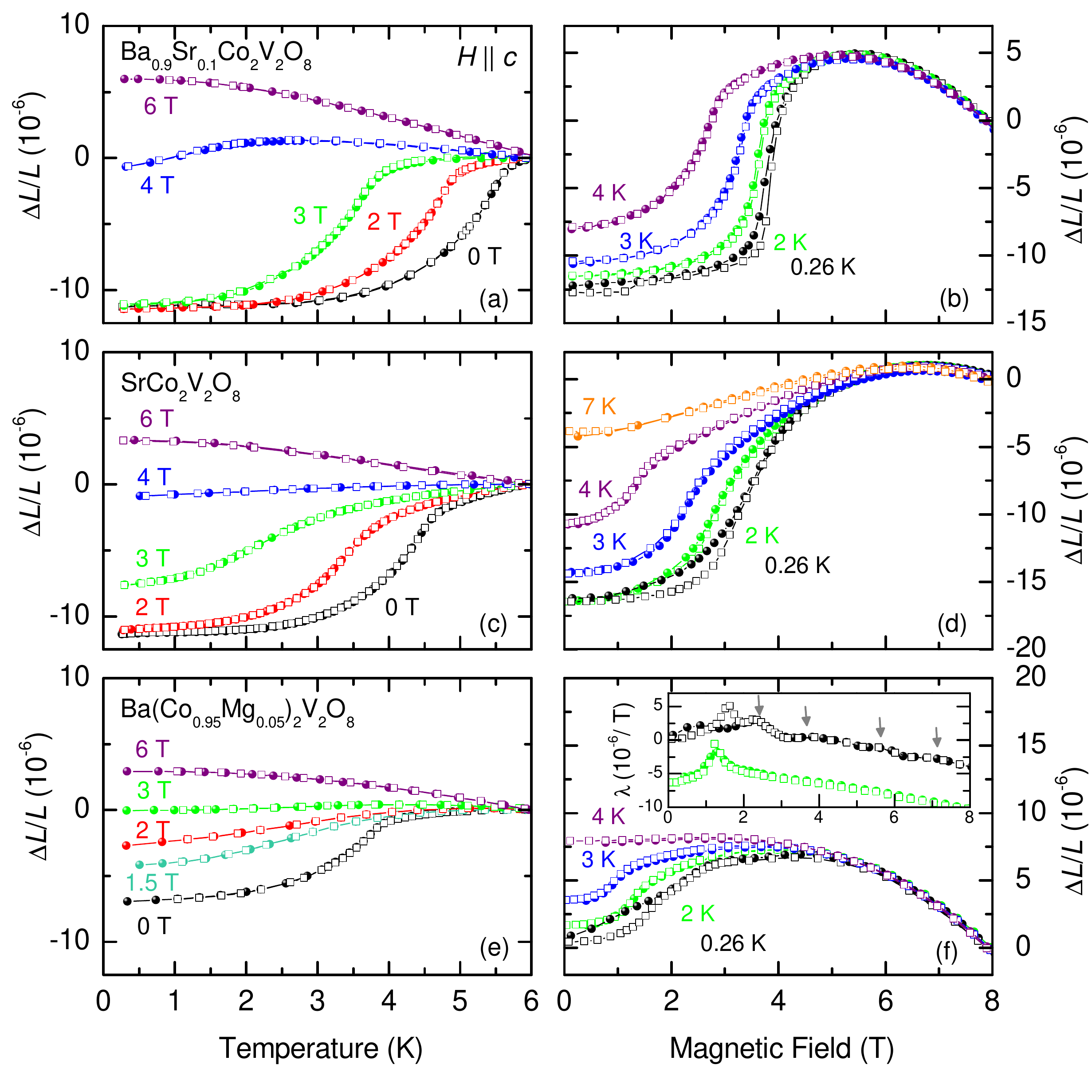}
	\caption{(color online) Thermal-expansion (left) and magnetostriction (right) $\Delta L(T,H)/L$ measured along the $c$ axes of \basrA , \srco , and  Ba(Co$_{0.95}$Mg$_{0.05}$)$_2$V$_2$O$_8$ (from top to bottom) for magnetic fields $H\|c$. Open and closed symbols are measured with increasing and decreasing temperature  or magnetic field, respectively. The inset displays the corresponding field derivatives $\lambda = 1/L\, \partial \Delta L / \partial H$ for \bacomg, which show additional features in the high-field low-temperature range as marked by the arrows.}
	\label{fig:tadHpc}
\end{figure}

Fig.~\ref{fig:tadHpc} displays representative thermal-expansion and magnetostriction  data of \basrA , \srco , and \bacomg measured along the $c$ axis for $H\|c$. In zero field, the antiferromagnetic ordering is accompanied by a spontaneous contraction, whose magnitude does not change with the Sr content, but is reduced by about a factor of 2 in the Mg-substituted crystal. On increasing the magnetic field, the transition temperature rapidly decreases and also the spontaneous contraction diminishes until it vanishes completely above a critical field of about 3 to 4~T, depending on the composition. Such a strong suppression of the N\'{e}el order in the field range up to 4~T is also present in the pure \baco , and above 4~T an incommensurate phase is observed below about 1~K~\cite{Kimura2008A,Grenier2013}. The corresponding temperature-dependent transitions from the paramagnetic to the IC phase are, however, very broad and only cause tiny anomalies in the thermal-expansion data; see Ref.~\onlinecite{Niesen2013}. 

Thus, the fact that there are no anomalies resolvable in the temperature-dependent data of Fig.~\ref{fig:tadHpc} does not exclude that IC phases might also be present in the substituted crystals in the respective high-field low-temperature regions. Some indirect evidence in favor of IC phases stems from the magnetic-field dependent measurements shown in the right panels of Fig.~\ref{fig:tadHpc}. From the zero-field \tn down to about 2~K, these data signal continuous transitions without hysteresis effects between the measurements obtained either with increasing or decreasing magnetic field, whereas below about 1~K there are such hysteresis effects indicating first-order phase transitions. This change from continuous to discontinuous transitions fits to the observations in \baco that the transitions between the N\'{e}el ordered phase and the paramagnetic phase are of 2$^{nd}$ order, while those from the N\'{e}el ordered to the IC phase are of 1$^{st}$ order. 

\begin{figure}[t]
	\centering
		\includegraphics[width=1.00\linewidth]{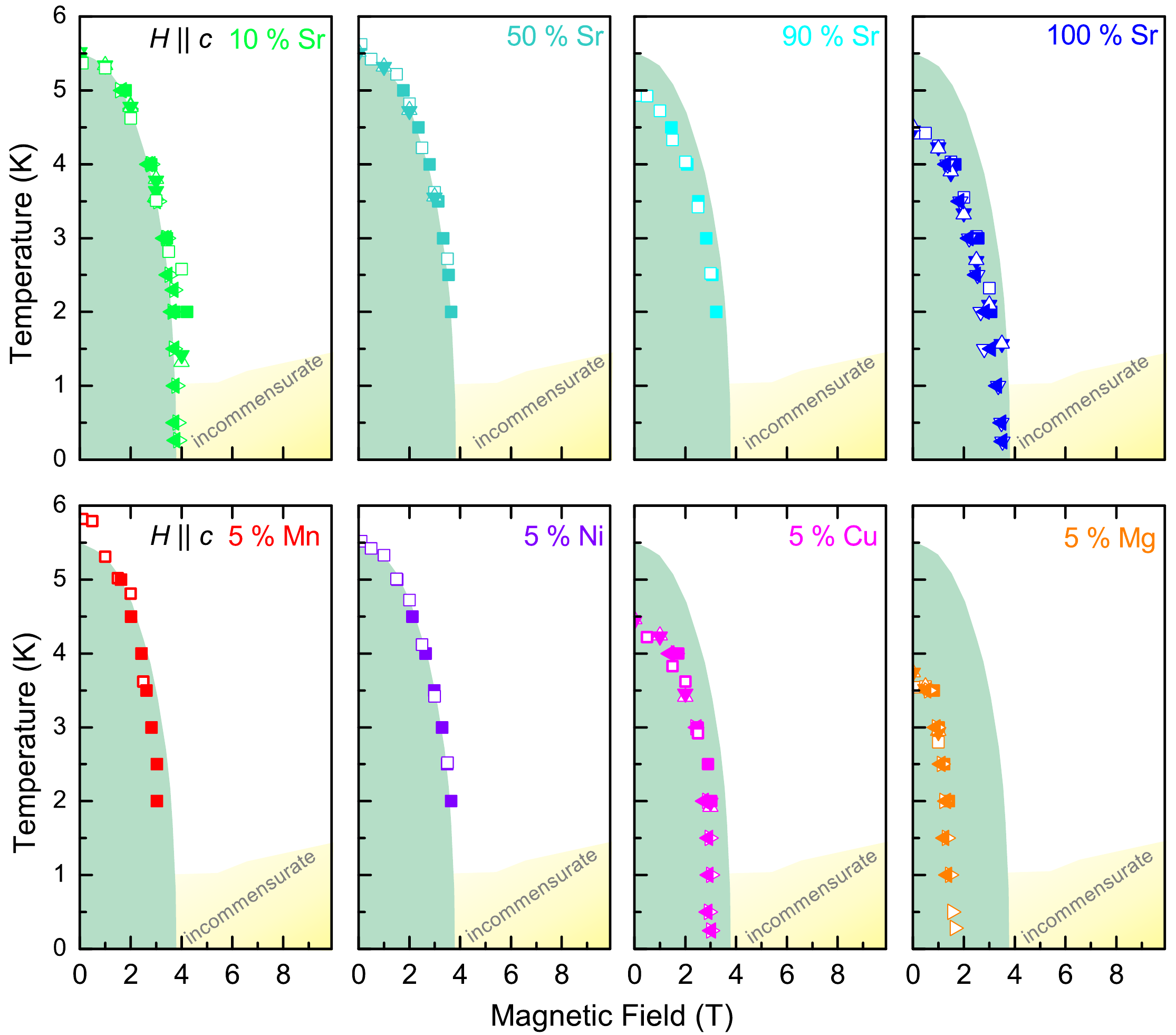}
	\caption{(color online) Temperature {\it vs.} magnetic-field phase diagrams of 
\basrco and \bacom for $H\|c$. For comparison, the N\'{e}el phase and the IC phase of the parent compound \baco (taken from Ref.~\onlinecite{Niesen2013}) are marked by shaded regions in green and yellow, respectively. The presence of IC phases in the substituted crystals needs further verification (see text).}
	\label{fig:phadiHpc}
\end{figure}

A particular feature is observed for the Mg-doped crystal. Here, the measurement at the lowest temperature $T=0.26$~K reveals a sequence of weak and rather broad anomalies, which are not seen at higher temperature; see inset of Fig.~\ref{fig:tadHpc}. In \baco, a continuous variation of the incommensurability $\Delta k(H)$ with increasing magnetic field has been observed~\cite{Kimura2008A}. In analogy to the incommensurate phase of Spin-Peierls systems, this may be visualized by a continuous increase of the density of soliton/anti-soliton kinks (or domain walls), which as a function of the magnetic field have to be continuously rearranged in order to keep them equidistant~\cite{Bak1982,Buzdin1983,Lorenz1998,Kiryukhin1996}. 
Such a rearrangement, and thus the continuous variation of $\Delta k(H)$, will be significantly disturbed if the spin chains are effectively cut into finite chain segments, as it is the case due to the substitution of Co$^{2+}$ by non-magnetic Mg$^{2+}$ ions. Thus, we interpret the additional anomalies in the low-temperature magnetostriction data of \bacomg as an indication for discontinuous changes in the field dependence of $\Delta k(H)$. Note, however, that these anomalies are very weak and can only be seen in the Mg-substituted material and the fact that no such features are resolved for the other substitutions corresponds to a less effective cutting of the chains, in agreement, with the weaker influence on \tn . Nevertheless, there is a certain amount of disorder in all substituted materials, which disturbs the evolution of a well-defined incommensurability in the low-temperature high-field range. This may also explain why in the corresponding temperature-dependent measurements $\Delta L_c(T, H \|c>4\,{\rm T})$ no resolvable anomalies appear, which should be present if there were well-defined transitions to the IC phase. This observation was already discussed in our recent work on the pure \baco, where the transitions to the IC phase are very broad but still resolvable~\cite{Niesen2013}. In this context, it is also worth to mention that the \baco crystal studied in Ref.~\onlinecite{Niesen2013} was grown by spontaneous nucleation, whereas all the crystals used for the present work were grown by the image-furnace technique and none of them (including a \baco crystal) reveals resolvable anomalies of $\Delta L_c(T, H \|c>4\,{\rm T})$.

In Fig.~\ref{fig:phadiHpc} we summarize the temperature {\it vs.} magnetic-field phase diagrams for $H\| c$ of both substitution series. For comparison, the phase diagram of the pure \baco from  Ref.~\onlinecite{Niesen2013} is added by shaded regions of different colours  corresponding to the N\'eel phase and the IC phase. As can be expected from the rather weak decrease of the zero-field \tn, the phase diagrams of all compositions are very similar with respect to the N\'eel phase, in particular, the zero-temperature extrapolation of the critical field $H_c(T\rightarrow 0)$ changes rather weakly, see Table~\ref{tab:crit}. Concerning the IC phase, our data can only yield rather indirect evidence for its occurrence in the substituted materials and this question needs further investigation by diffraction techniques.

\subsection{Magnetic field perpendicular to [001]}

\begin{figure}[b]
	\centering
		\includegraphics[width=1.0\linewidth]{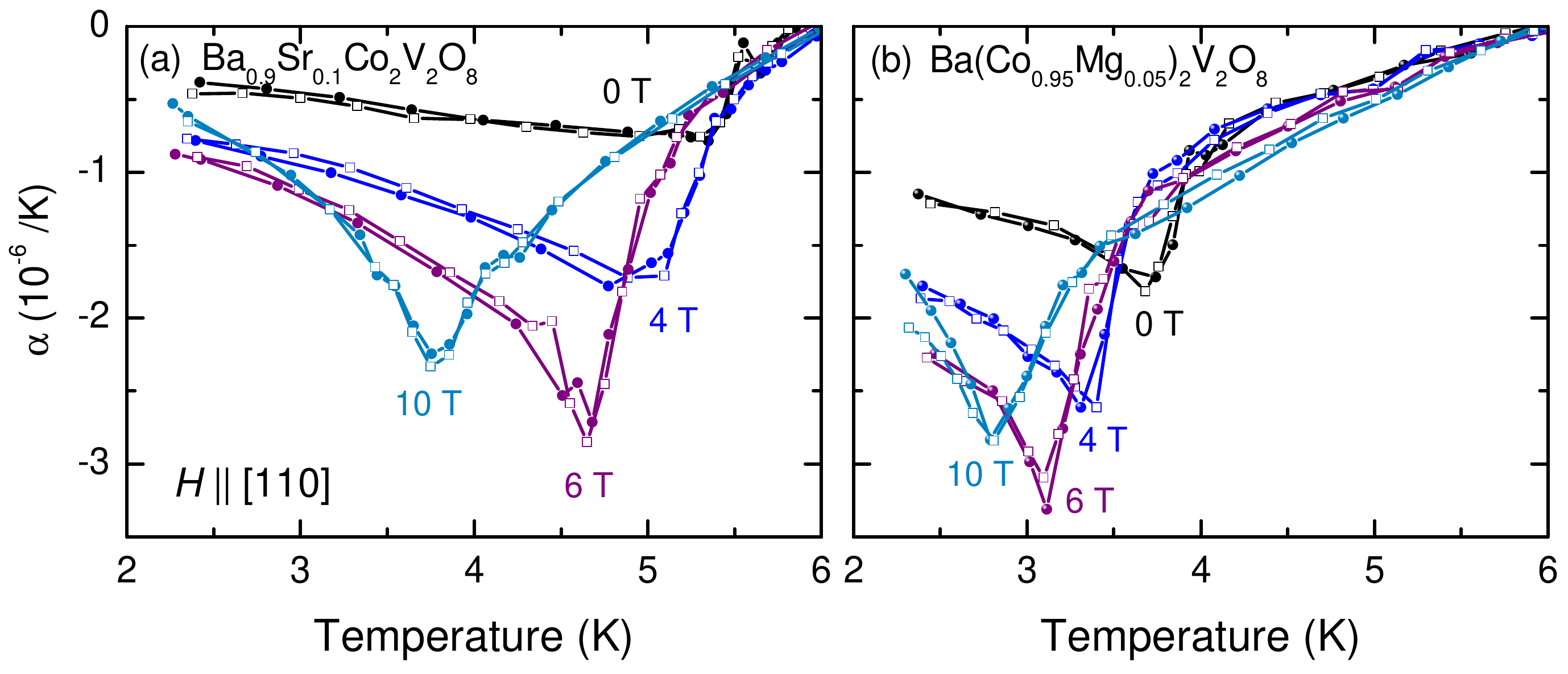}
	\caption{(color online) Representative thermal-expansion coefficients $\alpha=1/L_0\,\partial \Delta L/\partial T$  of (a) \basrA and (b) \bacomg for $H\|L\| [110]$ up to 10~T. Open and closed symbols are measured with increasing and decreasing temperature, respectively.}
	\label{fig:TAD-110}
\end{figure}

\begin{figure}[t]
	\centering		\includegraphics[width=1.00\linewidth]{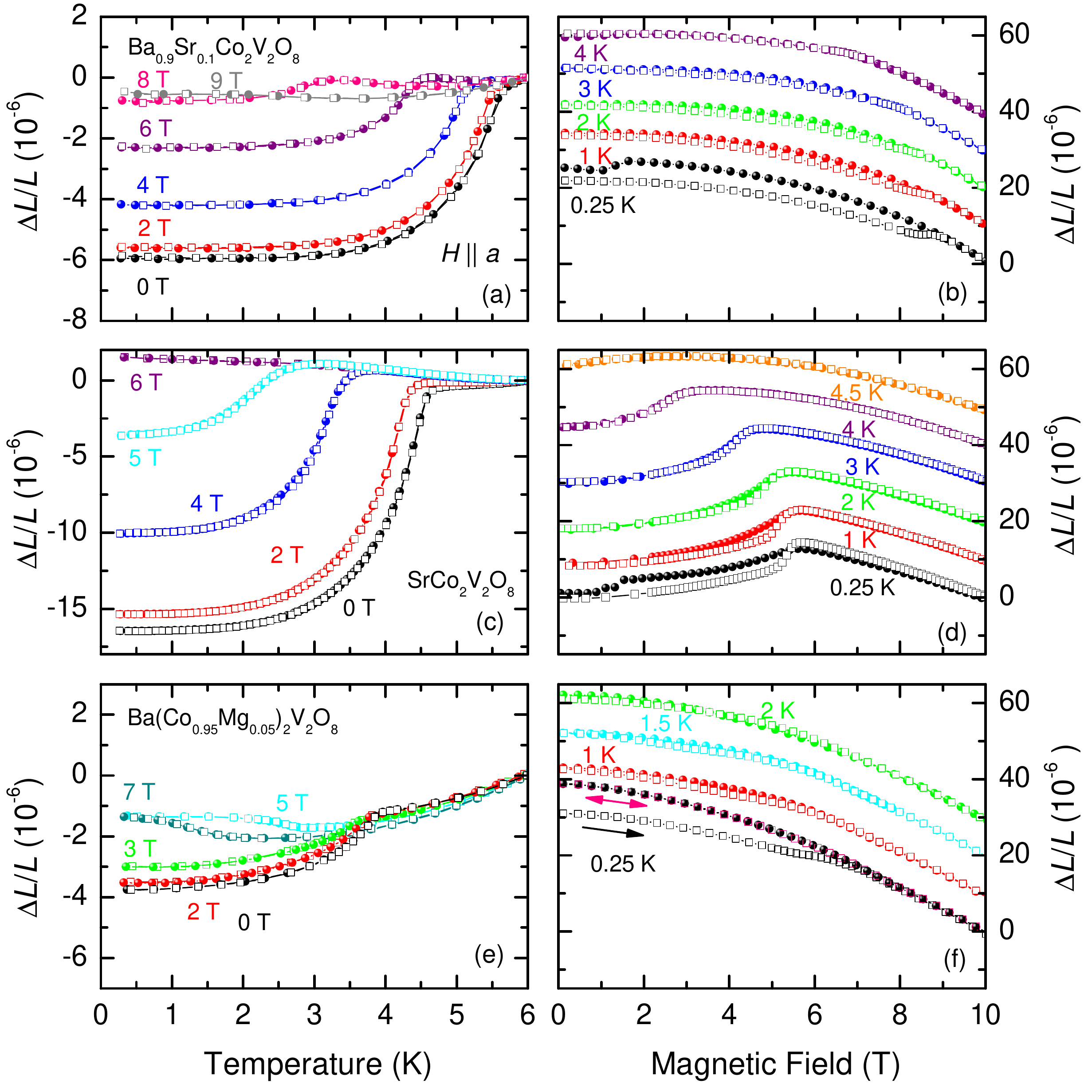}
	\caption{(color online) Thermal-expansion (left) and magnetostriction (right) $\Delta L(T,H)/L$ measured along the tetragonal $a$ axes of \basrA , \srco , and  Ba(Co$_{0.95}$Mg$_{0.05}$)$_2$V$_2$O$_8$ (from top to bottom) for $H\|a$. Open and closed symbols are measured with increasing and decreasing temperature or field, respectively. For clarity, the $\Delta L(H)/L$ curves measured at different constant temperatures are offset by $10^{-5}$ with respect to each others. The arrows in (f) mark a virgin curve obtained with increasing field after cooling the crystal in zero field, which differs from the subsequent $\Delta L(H)/L$ measured with decreasing or increasing field.}
	\label{fig:tadHpa}
\end{figure} 

The observation of a pronounced in-plane anisotropy for magnetic fields applied perpendicular to the Ising axis [001] of \baco was a main new finding of Refs.~\onlinecite{Kimura2013, Niesen2013}. Thus, the question arises whether this in-plane anisotropy is preserved under (partial) substitution of Ba or Co.  Fig.~\ref{fig:TAD-110} displays representative thermal-expansion data $\alpha =\frac{1}{L}\frac{\partial \Delta L}{\partial T}$ of \basrA and \bacomg, which were measured for  $H\| [110]$ up to 10~T and in both cases the decrease $\partial T_{\rm N}/ \partial H \approx -0.1$~K/T is rather weak.

\begin{figure}[t]
	\centering
		\includegraphics[width=1.00\linewidth]{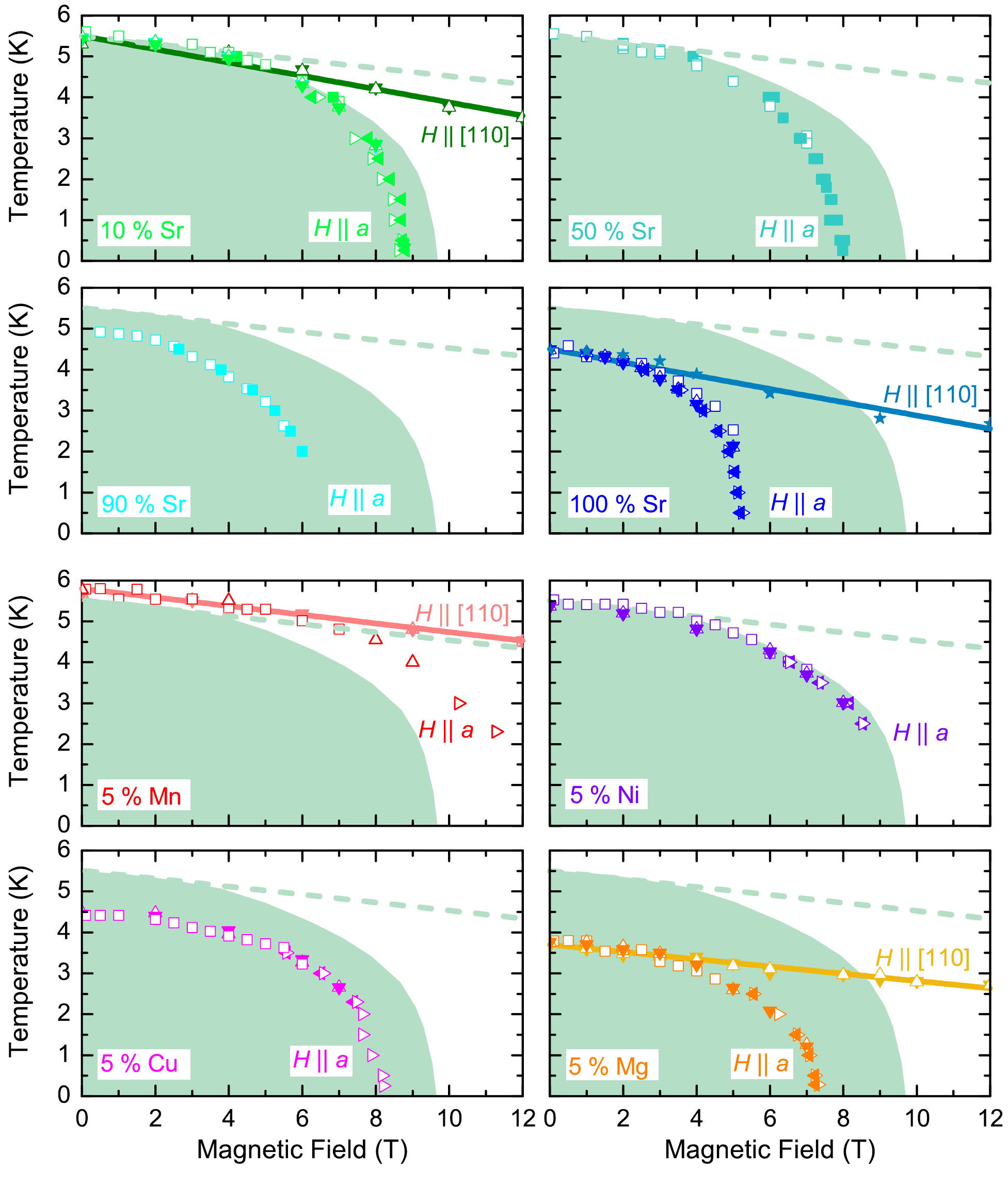}
	\caption{(color online) Temperature {\it vs.} magnetic-field phase diagrams of 
\basrco and \bacom for magnetic fields $H\|a$ (symbols). The N\'{e}el phase of the parent compound \baco is marked by the shaded region. The solid lines in the respective colours mark the corresponding phase boundaries $T_{\rm N}$ for $H\| [110]$, which are essentially linear in the studied field range (the dashed line corresponds to the parent compound).}
	\label{fig:phadiHpa}
\end{figure}
 
In contrast, a much stronger suppression of the N\'eel phase is observed for $H\| a$ as is shown in Fig.~\ref{fig:tadHpa}. The $\Delta L(T)$ data reveal that with increasing magnetic field \tn rapidly decreases and vanishes completely in the field range between about 5 and 9~T, depending on the composition. The spontaneous strains of the $a$ axes in the N\'eel phases are of different magnitudes and of different signs for the different crystals (see also Fig.~\ref{fig:tadnull}) and, in some cases, the spontaneous strain even changes from a contraction to an elongation at a certain intermediate field, see Fig.~\ref{fig:tadHpa}(e). As discussed in detail in Ref.~\onlinecite{Niesen2013}, these effects are related to the orthorhombic twin domains and a partial magnetic-field induced detwinning. Such detwinning effects also explain the hysteresis between the $\Delta L(H)$ curves obtained with increasing field after zero-field cooling and those measured either with decreasing field or in the subsequent field-increasing run, as is marked by the arrows in Fig.~\ref{fig:tadHpa}(f).

The resulting phase diagrams for $H\|a$ for all samples are summarized in Fig.~\ref{fig:phadiHpa} and again the corresponding phase diagram of the pure \baco is shown as the shaded area for comparison. In addition, for some of the crystals we have also added the weakly field-dependent $T_{\rm N}(H || [110])$ phase boundaries, which are shown as solid lines. The corresponding phase boundary of \baco is depicted by a dashed line. This comparison clearly shows that the pronounced in-plane anisotropy remains present over the entire substitution range studied here. This finding, together with the phase diagrams for $H\| c$, and the characteristic detwinning effects under uniaxial pressure or magnetic fields along $a$ strongly suggest that the basic features of the magnetic structure of \baco are neither changed by the complete substitution of Ba by Sr nor by the partial substitutions of Co by other divalent ions. In particular, our findings rule out the previous suggestion that the magnetic structures of \baco and of \srco were fundamentally different~\cite{He2006Sr,He2007Sr}. Apparently, the authors of Ref.~\onlinecite{He2007Sr} were misled by the fact that the in-plane anisotropy was not known and that they accidentally compared the strongly field-dependent phase boundary of \srco for $H\| a$ with the weakly field-dependent one of \baco for $H\| [110]$.

\begin{figure}[t]
	\centering
		\includegraphics[width=1.00\linewidth]{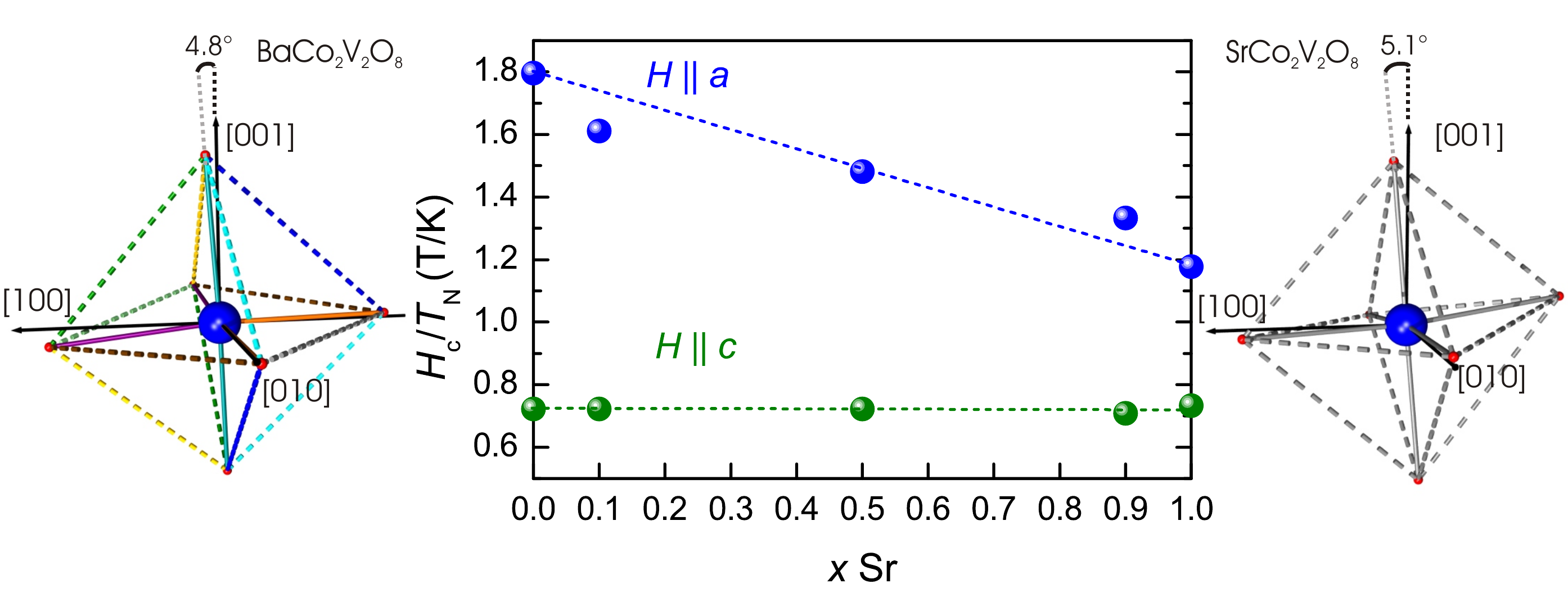}
		\caption{(color online) Comparison of the ratios  $H_{crit}^{\| i}(T\rightarrow 0)/T_{\rm N}(H=0)$ for applied magnetic fields $H\| a$ and  $H\| c$. The distorted CoO$_6$ octahedra of the end members \baco (equal Co-O or O-O bond lengths are given in the same colors) and \srco (all bond lengths are different~\cite{Note2}) based on Refs.~\onlinecite{Wichmann86, Osterloh94}.}
	\label{fig:rel}
\end{figure}

For the \basrco series, we find that with increasing Sr content the zero-temperature extrapolation $H_{crit}^{\| a}(T\rightarrow 0)$ decreases significantly faster than the corresponding zero-field \tn. Fig.~\ref{fig:rel} compares the ratios $H_{crit}^{\| i}(T\rightarrow 0)/T_{\rm N}(H=0)$ for the different field directions as a function of the Sr content. For $H\|c$ there is practically no change, while $H_{crit}^{\| a}(T\rightarrow 0)/T_{\rm N}(H=0)$ strongly decreases with $x$. As is argued in Ref.~\onlinecite{Kimura2013}, the magnitude of $H_{crit}^{\| a}$ is related to a staggered $g$ tensor, which itself results from a tilt of the local quantization axes of the Co ions with respect to the $c$ axis. Thus, the systematic decrease of $H_{crit}^{\| a}(T\rightarrow 0)/T_{\rm N}(H=0)$  with increasing Sr content would be naturally explained if the local quantization axes are continuously tilted away from the $c$ axis. Comparing the refined structural data~\cite{Wichmann86, Osterloh94,Note2} of the two end members of \basrco supports this interpretation. Fig.~\ref{fig:rel} also displays the corresponding CoO$_6$ octahedra, which for \baco consist of 3 pairs of equal Co--O and 5 pairs of equal O--O bond lengths, whereas such pair-wise equal bond lengths are not present in \srco. Thus, the CoO$_6$ octahedra of \srco are significantly more distorted than those of \baco and, moreover, the tilt angle of the short upper apical Co-O bond increases from $\simeq 4.8^\circ$ to $\simeq 5.1^\circ$. These observations support the above idea that by increasing the Sr content the tilt angle of the local quantization axes, and thus the staggered $g$ factor, increases with increasing Sr content. Consequently, an external field $H\|a$ would cause larger staggered fields $h_b$ and result in a systematic decrease of $H_{crit}^{\| a}(T\rightarrow 0)/T_{\rm N}(H=0)$ with increasing $x$. 

\section{Summary}

By studying the substitution series \basrx ($0 \leq x \leq 1$) and \bacom ($M=$ Mn, Ni, Cu, Mg), we have investigated the influence of out-of-chain and in-chain substitutions on the magnetic properties of the quasi-1D effective Ising spin-1/2 chain material \baco. Concerning the paramagnetic phase, we find that the strong Ising anisotropy of the magnetic susceptibility remains almost unchanged, that is, the pronounced easy-axis anisotropy of the Co$^{2+}$ spins due to the crystal electric field is essentially preserved. Moreover, we observe rather weak (typically less than $\simeq 1$~K) changes of the zero-field N\'{e}el temperatures \tn  for almost all substitutions. A significantly larger decrease of \tn is induced by the partial substitution of Co$^{2+}$ by non-magnetic Mg$^{2+}$, which can be traced back to an effective cutting of the 1D chains into finite chain segments. This segmentation is less effective, but still moderate for the partial substitution by Cu$^{2+}$, while almost no effect is observed in the partially Ni$^{2+}$- or Mn$^{2+}$-substituted materials. Thus, it appears that the in-plane substitutions affect the NN intra-chain coupling, while the Sr substitution  seems to mainly influence the interchain interactions $J^\perp_i$. A significant frustration of $J^\perp_{NN}$ and $J^\perp_{NNN}$ naturally explains that the magnetic structure of \baco is lower than tetragonal and results in two types of magnetic domains, which are coupled to structural domains via magnetoelastic coupling. The  twinning ratio of these domains can be influenced by uniaxial pressure (or magnetic field), what is clearly seen in high-resolution data of the macroscopic thermal expansion. This very characteristic feature of the magnetic structure of \baco is also observed over the full range of Sr substitution as well as for various in-chain substitutions. This gives clear evidence that the magnetic structure is essentially  preserved for all these substitutions and contradicts earlier assumptions~\cite{He2006Sr,He2007Sr} of a different magnetic structure in \srco. We suspect that the weakly decreasing \tn with increasing Sr content arises from an increasing frustration of the interchain couplings.

Concerning the temperature $vs.$ magnetic-field phase diagrams, we also find that all the basic features are essentially preserved for the different kinds of substitutions. For $H\|c$, the region of the antiferromagnetic phase roughly scales with the magnitude of the zero-field \tn. Above critical fields $H_{crit}^{\| c}$ ranging from about 2 to 4~T, our data yield some indirect evidence for the presence of incommensurate phases, but this latter point needs further verification by diffraction techniques. For $H\perp c$, we observe that the pronounced in-plane anisotropy between $H\| [110]$ and $H\| [100]$, which has been detected only recently in the parent compound \baco ~\cite{Kimura2013,Niesen2013}, is also present in  the substituted materials. In particular, we find that this in-plane anisotropy systematically increases with increasing Sr content. Following the interpretation of Ref.~\onlinecite{Kimura2013} that the in-plane anisotropy results from staggered $g$ tensors due to periodically tilted local quantization axes, an increasing tilt angle would naturally explain the enhanced in-plane anisotropy for the Sr-substituted samples. As an outlook, we would like to emphasize that most of our conclusions are based on macroscopic quantities and various aspects should therefore be studied in more detail by microscopic techniques. In \baco, for example, the proposed staggered fields for $H\perp c$ should be verified by the observation of transverse staggered moments, which differ for large magnetic fields applied either along [100] or [110]. Moreover, the magnetic structure of \srco remains to be finally clarified in zero-field as well as in high fields applied along [100], [110], and [001].

\begin{acknowledgments}
We acknowledge helpful discussions with M.~Braden, B.~Grenier, M.~Garst, M.~Gr\"{u}ninger, G.~Kolland, and N.~Qureshi. This work has been supported by the Deutsche Forschungsgemeinschaft via SFB 608 and through the Institutional Strategy of the University of Cologne within the German Excellence Initiative.
\end{acknowledgments}


\end{document}